\documentclass[10pt]{iopart}
\usepackage{iopams}
\usepackage{color,hyperref,graphicx}

\definecolor{orange}{rgb}{0.8,0.6,0.1}

\newtheorem{proposition}{Proposition}

\newtheorem{definition}{Definition}

\begin{document}

\title[Convex ordering and quantification of quantumness]{Convex ordering and quantification of quantumness}

\author{J Sperling}
\address{Arbeitsgruppe Theoretische Quantenoptik, Institut f\"ur Physik, Universit\"at Rostock, D-18051 Rostock, Germany}
\ead{jan.sperling@uni-rostock.de}
\author{W Vogel}
\address{Arbeitsgruppe Theoretische Quantenoptik, Institut f\"ur Physik, Universit\"at Rostock, D-18051 Rostock, Germany}
\ead{werner.vogel@uni-rostock.de}

\begin{abstract}
	The characterization of physical systems requires a comprehensive understanding of quantum effects.
	One aspect is a proper quantification of the strength of such quantum phenomena.
	Here, a general convex ordering of quantum states will be introduced which is based on the algebraic definition of classical states.
	This definition resolves the ambiguity of the quantumness quantification using topological distance measures.
	Classical operations on quantum states will be considered to further generalize the ordering prescription.
	Our technique can be used for a natural and unambiguous quantification of general quantum properties whose classical reference has a convex structure.
	We apply this method to typical scenarios in quantum optics and quantum information theory to study measures which are based on the fundamental quantum superposition principle.
\end{abstract}


\begin{indented}
	\item[]\today,\submitto{\PS}
\end{indented}

\vspace{2pc}

\maketitle
\ioptwocol

\section{Introduction}

	Characterizing the differences between the quantum and classical domain of physics is of fundamental interest for uncovering the quantumness of nature.
	Typically there are quantum counterparts to classical physics, such as coherent states in the system of the harmonic oscillator, or product states in the field of compound systems.
	Using classical statistical mixing, these pure states may be generalized to mixed classical ones.
	Thus, we obtain convex sets of states having a classical analogue with respect to a given physical property.
	Different measures have been introduced for quantifying the amount of quantumness of states having no such classical correspondence.
	These measures induce an ordering prescription enabling us to compare the quantumness of different states.

	In the system of the harmonic oscillator, one of the early attempts to quantify the amount of nonclassicality has been given by the trace-distance of an arbitrary state to the set of all classical ones being mixtures of coherent states~\cite{TraceDist,TraceDist2}.
	This led to a number of distance based nonclassicality probes, e.g., Hilbert-Schmidt-norm~\cite{HSDist,HSDist2} or the Bures distance~\cite{BuresDist} measures.
	Some nonclassicality metrics are based on the amount of Gaussian noise which is needed for the elimination of any quantum interference within the corresponding phase-space representation~\cite{Gauss1,Gauss12,Gauss2} or they directly use the negativities within the quasiprobability distribution~\cite{NegWiegner,FM10} as an indicator of the amount of nonclassicality.

	Another method for the quantification of nonclassicality is given via the potential of a state to generate entanglement~\cite{EntPot}.
	This translates quantumness of a single-mode harmonic oscillator to the quantification of entanglement.
	The axiomatic definition of general entanglement measure is given in~\cite{AxiomEntM,AxiomEntM2,AxiomEntM3}.
	This definition is based on so-called local operations and classical communications mapping separable quantum states onto separable ones.
	Under all examples of entanglement measures, there is one which is of particular interest for our considerations: the Schmidt number~\cite{SchmNummer,TerhalSN}.
	It has been shown that this entanglement measure has some advantageous properties in relation to other measures~\cite{SchmidtUni}.
	In particular, the degree of nonclassicality of a single mode system is directly transformed into the same Schmidt number using linear optics~\cite{UniQuant}.

	For some applications not all states with the same amount of quantumness are equally useful.
	For example, it has been shown that states can be too entangled for quantum computation~\cite{Eisert-QC}.
	Consequently, operational nonclassicality and entanglement measures have been introduced~\cite{SchmidtUni,Gehrke}.
	In particular quantum information protocols require information related measures of quantum effects.
	For example, the Fischer information~\cite{Fischer,FKMMSV10} is such a proper operational probe.
	More generally, entropic measures have been intensively studied~\cite{EntropyInformation,Context}.
	It has been shown that entropic inequalities and tomographic information can determine quantum correlations~\cite{MM09, MM14a,MM14}.
	In general, a given operational, distance-based, or entropic metrics induces an ordering prescription, which yields a particular sorting of quantum states regarding their amount of quantumness for some applications.

	In the current contribution we will use the inverse approach, i.e.: a convex ordering prescription of quantum states will imply a canonic measure.
	It will be shown that distance measures are, in general, not completely suitable for ordering quantum states unambiguously.
	Studying the algebraic implications of the definition of convex sets, we rigorously formulate an ordering procedure which does not depend on a distinct topological distance.
	We expand this method to include classical operations being especially defined for a particular notion of quantumness under study.
	The obtained sorting procedure induces a corresponding quantumness measures in a natural way.
	We apply this method to basic examples, such as entanglement, nonclassicality, and quantum information, showing the importance of the quantum superposition principle for the quantification of different quantum features.

	The paper is structured as follows.
	In Sec.~\ref{Sec:Motivation} we motivate our treatment.
	An unambiguous convex ordering prescription will be proposed in Section~\ref{Sec:Ordering}.
	In Sec.~\ref{Sec:ClassicalOperations} we include classical operations to further enhance the ordering technique.
	We introduce an axiomatic quantification and we study measures that count quantum superpositions in Sec.~\ref{Sec:Quantification}.
	A summary and conclusions are given in Sec.~\ref{Sec:SumCon}.

\section{Motivation}\label{Sec:Motivation}
	Let us consider the convex set of all (pure and mixed) quantum states, $\mathcal Q$, and a closed, non-empty, and convex subset $\mathcal C\subset\mathcal Q$.
	The elements of $\mathcal C$ are supposed to be states with a given classical property, e.g.: separable states, $\mathcal C_{\rm sep}={\rm conv}\{|a\rangle\langle a|\otimes|b\rangle\langle b|:|a\rangle\in\mathcal H_A \wedge |b\rangle\in\mathcal H_B\}$, or coherent states, $\mathcal C_{\rm coh}={\rm conv}\{|\alpha\rangle\langle \alpha|:\alpha\in\mathbb C\}$.
	The general task is the determination of the amount of quantumness of an arbitrary quantum state $\rho\in\mathcal Q$ with respect to the classical property under study.

	The convexity of the set $\mathcal C$ guarantees that a mixing of two classical states remains classical.
	This is important, because it ensures that statistical averaging cannot increase quantum correlations.
	The closure of $\mathcal C$ is motivated by the argument that a convergent sequence of classical states should have its limit in the classical domain too.
	These fundamental requirements ensure that a classical system remains classical employing classical operations and classical statistics.
	Let us note that the property of quantum discord does not meet these conditions, since a non-zero discord can be obtained from a classical mixing of two zero discord states~\cite{DiscordReview}.

\begin{figure}[h]
	 \centering
	 \includegraphics[width=5cm]{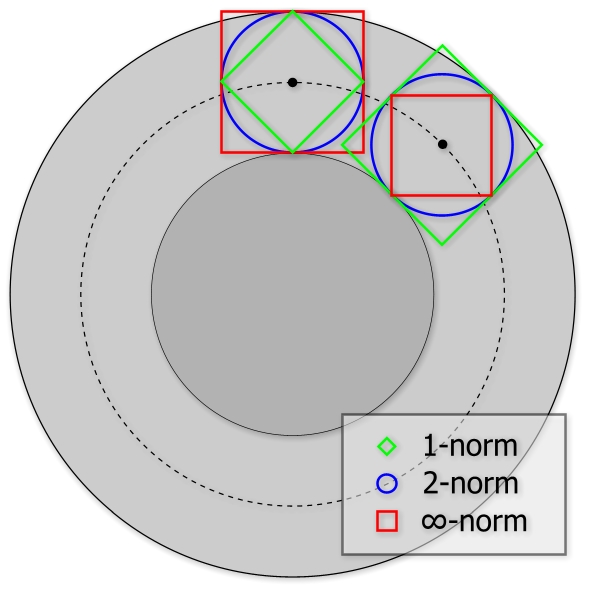}
	 \caption{(color online)
	 	The dark gray area represents $\mathcal C$, and both gray areas depict $\mathcal Q$.
	 	The upper point represents $y_1$, the other one represents $y_2$.
	 	The blue circles are the spheres in $2$-norm showing the distance to $\mathcal C$.
	 	The equal size of them implies an equal $2$-norm-distance for both points.
	 	The green squares represent the spheres around the considered points in $1$-norm.
	 	In the case of the $1$-norm, the square around $y_2$ is larger than those around $y_1$.
	 	Whereas for the $\infty$-norm spheres (red squares) the relation is the other way around.
	 }
	 \label{Fig:Norms}
	 \end{figure}

	One way of ordering quantum states is given by the distance of these states to the set of classical states $\mathcal C$.
	Here we will show that sorting quantum states by a distance cannot lead to one distinct order of states.
	For the time being, let us assume a two dimensional convex set $\mathcal C$.
	Using an appropriate coordinate transformation, this classical set $\mathcal C$ can be assumed to be a sphere -- with respect to the Euclidean norm $\|\,\cdot\,\|_2$ -- in the form:
	\begin{eqnarray}
		\mathcal C=\{x\in\mathcal Q:\, \|x\|_2\leq 1/2\}.
	\end{eqnarray}
	Now we may choose two nonclassical elements $y_1,y_2\in\mathcal Q\setminus\mathcal C$, which are given in the standard basis: $y_1=(1,0)^{\rm T}$ and $y_2=\frac{1}{\sqrt{2}}(1,1)^{\rm T}$.
	The distance $d_p$ to the set of classical states $\mathcal C$ in $p$-norm is given by
	\begin{eqnarray}
		d_p(y,\mathcal C)=\inf_{x\in\mathcal C}\|y-x\|_p.
	\end{eqnarray}
	For all $p$-norms the minimal distance of $y_1$ and $y_2$ to the classical states is obtained for $x_1=\frac{1}{2}(1,0)^{\rm T}\in\mathcal C$ and $x_2=\frac{1}{2\sqrt 2}(1,1)^{\rm T}\in\mathcal C$, respectively, cf. Fig.~\ref{Fig:Norms}.
	Thus, we can calculate $d_p(y_1,\mathcal C)$ and $d_p(y_2,\mathcal C)$ for different values of $p$,
	\begin{eqnarray}
		d_p(y_1,\mathcal C)=\|y_1-x_1\|_p=\left[\left(\frac{1}{2}\right)^p+0^p\right]^{1/p}=\frac{1}{2}\\
		\nonumber d_p(y_2,\mathcal C)=\|y_2-x_2\|_p\\
		\phantom{d_p(y_2,\mathcal C)}
		=\left[\left(\frac{1}{2\sqrt 2}\right)^p+\left(\frac{1}{2\sqrt 2}\right)^p\right]^{1/p}
		=\frac{2^{1/p}}{2\sqrt 2}.
	\end{eqnarray}
	This result displays the paradox of the quantification of quantumness with distance measures in Fig.~\ref{Fig:Norms}.
	Depending on the choice of the norm, we can claim that: $y_1$ is more nonclassical than $y_2$ ($2<p\leq\infty$); or $y_1$ is less nonclassical than $y_2$ ($1\leq p<2$); or $y_1$ and $y_2$ have an equal nonclassicality ($p=2$).

	Let us note that this particular two-dimensional cut already provides the ambiguity of the distance-measure approach for any dimension of convex sets.
	Additionally, any monotonic function of a distance, for example entropies, will inherit this characteristic.
	While those metrics can be useful in an operational sense, they are not suitable for an unambiguous quantification of the quantumness property itself.
	In the following we will show that the convexity of the classical set serves as the key element to resolve this paradox.

\section{Ordering Quantum States}\label{Sec:Ordering}
	A convex set $\mathcal C$ is characterized through its algebraic definition,
	\begin{eqnarray}\label{Eq:Convexity}
		\rho,\rho'\in\mathcal C \wedge \lambda\in[0,1] \Rightarrow \lambda \rho+(1-\lambda)\rho'\in\mathcal C.
	\end{eqnarray}
	The question whether a general element $\rho\in\mathcal Q$ is in the convex set $\mathcal C$, or not, is independent of the choice of a distance.
	In addition, we show in~\ref{App:Normalization} that the normalization to ${\rm tr}\,\rho=1$ can be neglected from the mathematical point of view.
	For the quantification, we start with the formulation of a preorder relation $\preceq$.
	\begin{definition}\label{Def:Preorder}
		Two quantum states $\rho,\rho'\in\mathcal Q$ can be compared by $\preceq$:
		\begin{eqnarray*}
			\rho\preceq \rho' \Leftrightarrow \exists \gamma\in\mathcal C\, \exists \lambda\in[0,1]: \rho=\lambda \rho'+(1-\lambda)\gamma.
		\end{eqnarray*}
	\end{definition}
	This means a quantum state $\rho$ has less or equal nonclassicality compared with another state $\rho'$, if $\rho$ can be written as a classical statistical mixture of $\rho'$ and a classical state $\gamma$.
	Let us prove, that this relation fulfills the requirements of a preorder.
	\paragraph*{Proof.}
		$\preceq$ is reflexive: $\rho=1\rho+(1-1)\gamma\Rightarrow \rho\preceq \rho$; 
		$\preceq$ is transitive:  $\rho_1\preceq \rho_2$ and $\rho_2\preceq \rho_3$ imply
		\begin{eqnarray*}
			&\rho_1=\lambda \rho_2+(1-\lambda)\gamma_1 \wedge \rho_2=\kappa \rho_3+(1-\kappa)\gamma_2 \Rightarrow\\
			&\rho_1=\lambda\kappa \rho_3 +(1-\lambda\kappa)\gamma_3 \Rightarrow \rho_1\preceq \rho_3,
		\end{eqnarray*}
		with $\gamma_3=\frac{\lambda(1-\kappa)}{1-\lambda\kappa}\gamma_2+\frac{1-\lambda}{1-\lambda\kappa}\gamma_1$ and $\frac{\lambda(1-\kappa)}{1-\lambda\kappa}+\frac{1-\lambda}{1-\lambda\kappa}=1$.
		In conclusion, $\preceq$ is a preorder.
	\hfill$\blacksquare$

	For generating an order from the preorder $\preceq$, we consider the following equivalence~$\cong$.
	\begin{definition}\label{Def:Eqivalence}
	Two quantum states $\rho,\rho'\in\mathcal Q$ have the same order of quantumness, if
	\begin{eqnarray*}
		\rho\cong \rho' \Leftrightarrow \rho\preceq \rho' \wedge \rho'\preceq \rho.
	\end{eqnarray*}
	\end{definition}
	\paragraph*{Proof.}
		$\cong$ is reflexive: $\rho\preceq\rho\wedge\rho\preceq\rho$;
		$\cong$ is symmetric: $\rho\preceq\rho'\wedge\rho'\preceq\rho\Leftrightarrow\rho'\preceq\rho\wedge\rho\preceq\rho'$;
		$\cong$ is transitive: $\rho_1\cong\rho_2$ and $\rho_2\cong\rho_3$ are equivalent to
		\begin{eqnarray*}
			\rho_1\preceq\rho_2\quad\wedge\quad\rho_2\preceq\rho_1\quad\wedge\quad\rho_3\preceq\rho_2\quad\wedge\quad\rho_2\preceq\rho_3.
		\end{eqnarray*}
		Using the transitivity of $\preceq$, we obtain $\rho_1\preceq\rho_3\wedge\rho_3\preceq\rho_1$.
		Thus, $\cong$ is an equivalence relation.
	\hfill$\blacksquare$

	With respect to the equivalence $\cong$, the $\preceq$ preorder given in Definition~\ref{Def:Preorder} becomes an order.
	The missing property is that $\preceq$ must be antisymmetric,
	\begin{eqnarray}
		\rho\preceq \rho'\wedge\rho'\preceq \rho\Rightarrow\rho\cong\rho',
	\end{eqnarray}
	which is true, cf. Definition~\ref{Def:Eqivalence}.
	Thus, we have constructed a rigorous way to order quantum states.

	\begin{proposition}\label{Lem:MinClass}
	Classical states have a minimal and equal order, i.e.:
	\begin{eqnarray*}
		\gamma\in\mathcal C \wedge \rho\in\mathcal Q\Rightarrow \gamma\preceq\rho \mbox{ and }
		\gamma,\gamma'\in\mathcal C\Rightarrow\gamma\cong\gamma'.
	\end{eqnarray*}
	Any state $\rho\in\mathcal Q$ with a minimal order, $\rho\preceq\gamma\in\mathcal C$, is classical, $\rho\in\mathcal C$.
	\end{proposition}
	\paragraph*{Proof.}
		From $\gamma=0\rho+(1-0)\gamma$ and Definition~\ref{Def:Preorder} follows $\gamma\preceq\rho$.
		Hence we find for all classical states $\gamma,\gamma'\in\mathcal C$: $\gamma\preceq\gamma'\wedge\gamma'\preceq\gamma$; and therefore $\gamma\cong\gamma'$. 
		If $\rho\preceq\gamma\in\mathcal C$, i.e. $\exists\gamma'\in\mathcal C,\lambda\in[0,1]:\rho=\lambda\gamma+(1-\lambda)\gamma'$ , then $\rho$ is a convex combination of classical states and therefore classical.
	\hfill$\blacksquare$

\begin{figure}[h]
	 \centering
	 \includegraphics[width=5cm]{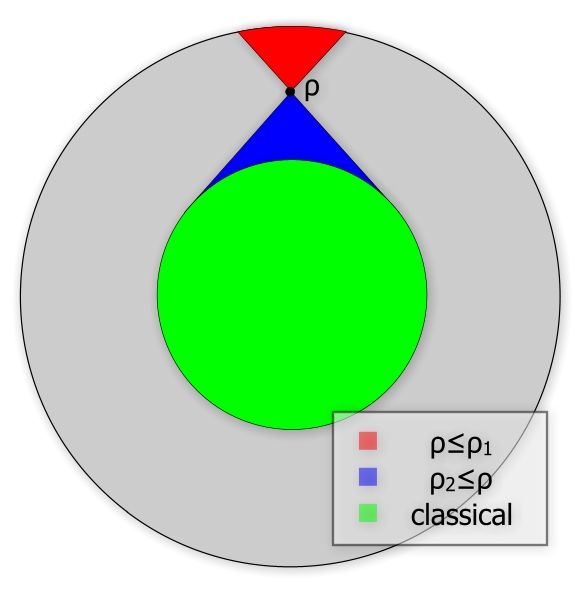}
	 \caption{(color online)
	 	The inner green area represents $\mathcal C$, and the complete area represents $\mathcal Q$.
	 	A nonclassical element $\rho$ is given.
	 	All elements $\rho_1$ in the red (triangular) area above $\rho$ fulfill: $\rho\preceq\rho_1$.
	 	All elements $\rho_2$ in the green and blue area below $\rho$ fulfill: $\rho_2\preceq\rho$.
	 }
	 \label{Fig:NonclOrder}
	 \end{figure}

	The Definitions~\ref{Def:Preorder}~and~\ref{Def:Eqivalence} provide an order of quantum states, which is solely based on the convex structure of $\mathcal C$.
	These definitions highlight the natural assumption that a statistical mixture of a nonclassical state with a classical one cannot become more nonclassical than the initial one, cf. Fig.~\ref{Fig:NonclOrder}.
	Further on, in Proposition~\ref{Lem:MinClass} it has been shown that this order implies that all classical states are the only ones with a minimal nonclassicality.
	The mixing property and the minimality property of classical states are essential for any quantification of nonclassicality.

\section{Classical Operations}\label{Sec:ClassicalOperations}
	A classical quantum state may evolves in an experiment or it propagates in a classical channel including noise effects.
	Thus we have to deal with operations which map our state within the set $\mathcal Q$.
	Operations with a classical counterpart must not increase the amount of quantumness.
	Therefore, we study transformations mapping classical states onto each other.
	\begin{definition}
		We call a linear operation $\Lambda:\mathcal Q\to\mathcal Q$ a classical one, if $\forall\gamma\in\mathcal C: \Lambda(\gamma)\in\mathcal C$.
		The set of all classical operations $\Lambda$ is denoted as $\mathcal{CO}$.
	\end{definition}
	\begin{proposition}\label{Prop:SemiGroup}
		The set $\mathcal{CO}$ is convex and a semi-group.
	\end{proposition}
	\paragraph*{Proof.}
		The convexity follows from the linearity of the operation space together with the convexity of the set of classical states,
		\begin{eqnarray*}
			(\lambda\Lambda_1+(1-\lambda)\Lambda_2)(\gamma)=\lambda\underbrace{\Lambda_1(\gamma)}_{\in\mathcal C}+(1-\lambda)\underbrace{\Lambda_2(\gamma)}_{\in\mathcal C}\in\mathcal C.
		\end{eqnarray*}
		The semi-group property is  given by
		\begin{eqnarray*}
			\Lambda_1,\Lambda_2\in\mathcal{CO}: (\Lambda_1\circ\Lambda_2)(\gamma)=&\Lambda_1(\Lambda_2(\gamma))\in\mathcal C,
		\end{eqnarray*}
		with the identity ${\rm Id}(\gamma)=\gamma\in\mathcal C$, being classical.
	\hfill$\blacksquare$
	
	These classical operations or channels can be considered as quantum physical systems having a classical analogue.
	This includes interactions which evolve states in a classical way, or mix them with classical noise.
	For special quantum tasks it might be also useful to consider only sub-semi-groups of $\mathcal{CO}$, for example one-way classical communications for entanglement or phase rotations and phase dispersion for coherent states.
	
	Now, we have to verify that classical operations do not change the previously defined order.
	Therefore we formulate the following proposition.
	\begin{proposition}\label{Theo:ClassOp}
		{\rm (i)} A classical operation does not change the order, $\rho\preceq\rho' \Rightarrow \Lambda(\rho)\preceq\Lambda(\rho')$.
		{\rm (ii)} Mixing a quantum state with a classical one, is a classical operation.
	\end{proposition}
	\paragraph*{Proof.}
		For (i) let us consider two states with $\rho\preceq\rho'$ and a classical operation $\Lambda$, $\rho=\lambda\rho'+(1-\lambda)\gamma$, which implies $\Lambda(\rho)=\lambda\Lambda(\rho')+(1-\lambda)\Lambda(\gamma)$.
		Together with $\Lambda(\gamma)\in\mathcal C$ and Definition~\ref{Def:Preorder} we obtain (i).
		For claim (ii), we consider that a state $\rho$ is mixed with a classical state $\gamma$, $\lambda\in[0,1]$,
		\begin{eqnarray*}
			\rho'&=\lambda\rho+(1-\lambda)\gamma\\
			&=\lambda{\rm Id}(\rho)+(1-\lambda)({\rm tr}\rho)\gamma\\
			&=\left[\lambda{\rm Id}(\,\cdot\,)+(1-\lambda)({\rm tr}(\,\cdot\,))\gamma\right](\rho)=\Lambda(\rho).
		\end{eqnarray*}
		The identical transformation ${\rm Id}$ and $({\rm tr}(\,\cdot\,))\gamma$ are classical, i.e., $\forall \gamma' \in\mathcal C:({\rm tr}\,\gamma')\gamma\in\mathcal C$,  and the convex structure of $\mathcal{CO}$ implies that $\Lambda\in\mathcal{CO}$.
	\hfill$\blacksquare$
	
	This means that classical operations are compatible with the order $\preceq$, and they cannot increase the quantumness of the initial state.
	Therefore, the order given in Definition~\ref{Def:Preorder} can be generalized by using Proposition~\ref{Theo:ClassOp}.
	\begin{definition}\label{Def:NO}
		A quantum state $\rho$ has a lower or equal order of nonclassicality than the state $\rho'$, $\rho\preceq\rho'$, iff $\exists\Lambda\in\mathcal{CO}:\rho=\Lambda(\rho').$
		They have the same nonclassicality, $\rho\cong\rho'$, if $\rho\preceq\rho'\wedge\rho'\preceq\rho$.
	\end{definition}
	This quantumness ordering prescription naturally generalizes the previous convex ordering with respect to $\mathcal C$ by including classical operations $\mathcal{CO}$.
	Condition~(ii) in Proposition~\ref{Theo:ClassOp} proves that the ordering includes the previous Definition~\ref{Def:Preorder}.
	In addition, the Definition~\ref{Def:NO} implies that all quantum states below a given state $\rho$ can be written as $\Lambda(\rho)$ for a classical operation $\Lambda$.
	Therefore it simply follows
	\begin{eqnarray}
		\Lambda(\rho)\preceq\rho.
	\end{eqnarray}
	Let us stress again, that the minimal states are uniquely classical ones.

	Now we want to further study properties of classical operations.
	A subgroup of $\mathcal{CO}$ are classical invertible maps $\mathcal{CO}_{-1}$, defined by
	\begin{eqnarray}
		\Lambda\in\mathcal{CO}_{-1}\Leftrightarrow \Lambda\in\mathcal{CO}\wedge\exists \Lambda^{-1}\in\mathcal{CO}.
	\end{eqnarray}
	These are classical operations which can be reversed, and the inverse is again a classical operation.
	This group always exists, since the identical transformation is its own inverse, ${\rm Id}\in\mathcal{CO}_{-1}$.
	The importance of this group is that it yields classes of quantum states with an equivalent order.
	Let us assume a classical invertible $\Lambda\in\mathcal{CO}_{-1}$ and an arbitrary state $\rho\in\mathcal Q$.
	It follows from $\rho'=\Lambda(\rho)$ that $\rho=\Lambda^{-1}(\rho')$.
	Together with the Definitions~\ref{Def:Eqivalence}~and~\ref{Def:NO}
	\begin{eqnarray}
		\rho\preceq\rho'\wedge\rho'\preceq\rho \Leftrightarrow \rho\cong\rho'.
	\end{eqnarray}
	Hence, it is possible to identify quantum states with an equal order of quantumness applying the group $\mathcal{CO}_{-1}$.
	\begin{proposition}
		All quantum states $\rho,\rho'\in\mathcal Q$, with $\rho'=\Lambda(\rho)$ and $\Lambda\in\mathcal{CO}_{-1}$, have an equal order of quantumness, $\rho\cong\rho'$.
		\hfill$\blacksquare$
	\end{proposition}
	Using the sphere shaped classical set in Fig.~\ref{Fig:NonclOrder}, we observe in this case that classical invertible maps are rotations around the center.
	This structure, in the generalized scenario, will lead subsequently to nested sets with increasing amount of quantum interferences.

\section{Axiomatic Quantification of Nonclassicality}\label{Sec:Quantification}
	So far, we have introduced the algebraic quantumness ordering prescription $\preceq$ on arbitrary classical, convex sets $\mathcal C$ that are closed under classical statistical mixtures and operations.
	Hence, a distance independent ordering technique is obtained.
	Eventually, we will use this approach to quantify the amount of quantumness in a natural way.

	Let us stress again that the standard approach is formulated in the opposite direction, i.e., a measure is proposed which implies an sorting of states. 
	Contrary, the approach under study starts from a convex geometric ordering.
	Using the derived ordering, we can properly define quantumness measures.
	This means that we can introduce functions $\mu$, which map a classical states, $\rho\in\mathcal C$, to a real number $\mu(\rho)$.
	\begin{definition}\label{Def:Measure}
		A function $\mu:\mathcal Q\to\mathbb R$ is a quantumness measure, if $\rho\preceq\rho'\Leftrightarrow \mu(\rho)\leq\mu(\rho')$.
	\end{definition}
	The definition says that the measure quantifies the ordering, which is given by the algebraic sorting $\preceq$.
	Since for all classical states $\gamma\in\mathcal C$ holds $\gamma\preceq\rho\in\mathcal Q$, we have $\mu(\rho)=\inf_{\gamma\in\mathcal C}\mu(\gamma)=:\mu_{\rm min}$ if and only if $\rho\in\mathcal C$, cf. Proposition~\ref{Lem:MinClass}.
	Typically, one uses the convention $\mu_{\rm min}=0$.
	From the definition also follows
	\begin{eqnarray}
		\mu(\rho)\geq\mu(\Lambda(\rho)),
	\end{eqnarray}
	for any classical operation $\Lambda\in\mathcal{CO}$.
	Moreover, equally ordered quantum states, $\rho\cong\rho'$, have an equivalent amount of quantumness,
	\begin{eqnarray}
		\rho\preceq\rho'\wedge\rho'\preceq\rho\,\Leftrightarrow\,\mu(\rho)\leq\mu(\rho')\wedge\mu(\rho')\leq\mu(\rho).
	\end{eqnarray}

	The here considered quantification of quantum states with nonclassical properties has been based only on the most elementary definition of statistical averaging (convexity of $\mathcal C$) and the physical need for classical transformations, $\mathcal{CO}$.
	We did not make any further assumption about the classical property itself.
	In the case of entanglement, Definition~\ref{Def:Measure} is equivalent to the axiomatic definition of entanglement measures~\cite{AxiomEntM,AxiomEntM2,AxiomEntM3} adding the compatibility with local invertible transformations.
	For nonclassicality in the notion of coherent states, Definition~\ref{Def:Measure} is equivalent to the algebraic approach in Refs.~\cite{UniQuant,Gehrke}.
	Note that the quantification procedure loses its generality if only subsets of $\mathcal{CO}$ are considered, as it is often done in entanglement theory by restricting the set of all separable operations to operational subset of so-called local operations and classical communication~\cite{RMP-Horo}.

\subsection{Quantumness measures based on the quantum superposition principle}\label{Sec:Example}
	As an example, we will consider in the following a measure which relies on the quantum superposition principle.
	Superpositions are the origin of the most fundamental differences between classical and quantum physics.
	Therefore, let us start with a set $\mathcal C_0$ of pure classical states, $|c\rangle\in\mathcal C_0$.
	The elements of the convex set $\mathcal C$ of all classical states are given by
	\begin{eqnarray}
		\gamma=\int_{\mathcal C_0} dP_{\rm cl}(c) |c\rangle\langle c|,
	\end{eqnarray}
	for a classical probability distribution $P_{\rm cl}$.
	Hence, a general classical state is a statistical mixtures of pure classical ones.
	For nonclassical states, $\rho\in\mathcal Q\setminus\mathcal C$, such a $P_{\rm cl}$ does not exist.
	The typical situation in quantum physics is that a generalized $P$ exists, but it has negativities.
	This scenario is relevant for the representations of both: expanding nonclassical states using coherent ones $\mathcal C_{0,\rm coh}=\{|\alpha\rangle:\, \alpha\in\mathbb C\}$ with the Glauber-Sudarshan representation~\cite{GSRep2,GSRep1}; and expanding entangled states by factorized ones $\mathcal C_{0,\rm sep}=\{|a\rangle\otimes|b\rangle:\,|a\rangle\in\mathcal H_A\wedge|b\rangle\in\mathcal H_B\}$ using optimized entanglement quasi-probabilities~\cite{EntRep}.

	Let us consider a classical operation, which has the following form,
	\begin{eqnarray}
		\Lambda(\rho)=M\rho M^\dagger,\mbox{ with } M|c\rangle=g(c)|f(c)\rangle,
	\end{eqnarray}
	with a classical valued function $f$, i.e. $|f(c)\rangle\in\mathcal C_0$, and a complex valued function $g$.
	This operation is a classical one,
	\begin{eqnarray}
		\nonumber\Lambda(\gamma)=\int_{\mathcal C_0} dP_{\rm cl}(c) M|c\rangle\langle c|M^\dagger\\
		\phantom{\Lambda(\gamma)}=\int_{\mathcal C_0} dP_{\rm cl}(c) |g(c)|^2|f(c)\rangle\langle f(c)|,
	\end{eqnarray}
	which is again (neglecting normalization, see~\ref{App:Normalization}) a statistical mixture of pure classical states.
	In case that $f$ is bijective and $g(c)\neq0$ for all $c$, we have a classical operation in $\mathcal{CO}_{-1}$,
	\begin{eqnarray}
		M^{-1}|c\rangle=\frac{1}{g(c)}|f^{-1}(c)\rangle.
	\end{eqnarray}
	Examples are local invertible maps $M=A\otimes B$ ($\exists A^{-1},B^{-1}$) for separable states, or, for coherent states,
	\begin{eqnarray}
		M=\exp[xa^\dagger a]\exp[ya]\exp[za^\dagger],
	\end{eqnarray}
	where $x,y,z\in\mathbb C$, the annihilation and creation operators $a$ and $a^\dagger$, respectively, and
	\begin{eqnarray}
		M|\alpha\rangle=\exp[xa^\dagger a]\exp[ya]\exp[za^\dagger]|\alpha\rangle\nonumber\\
		\phantom{M|\alpha\rangle}=e^{\frac{|z+\alpha|^2-|\alpha|^2}{2}+y(z+\alpha)}|(\alpha+z)e^{x}\rangle\in\mathcal C_0.
	\end{eqnarray}
	It is worth to note that the convex set of all classical operations, $\Lambda\in\mathcal{CO}$, can be written in the form of operator-sum decompositions~\cite{OpSumRep}, also called Krauss operators,
	\begin{eqnarray}\label{Eq:Krauss}
		\Lambda(\rho)=\sum_i M_i\rho M_i^\dagger.
	\end{eqnarray}

	Now we want to analyze a pure nonclassical state, which may be written as
	\begin{eqnarray}\label{Eq:SuperPosClassical}
		|\psi\rangle=\sum_{k=1}^r \psi_k |c_k\rangle,
	\end{eqnarray}
	with $|c_k\rangle\in\mathcal C_0$ and $r$ being the minimal number which allows this decomposition.
	This representation is possible for any pure state, if $\mathcal C_0$ includes at least a basis of the Hilbert space.
	Therefore, the state $|\psi\rangle$ is a superposition of $r$ classical states.
	The classical operator $M$ acts like
	\begin{eqnarray}
		M|\psi\rangle=\sum_{k=1}^r \psi_k g(c_k)|f(c_k)\rangle.
	\end{eqnarray}
	It is important that $M$ can only decrease the number $r$, for example, in the case $g(c_k)=0$ for some $k$ or for $f(c_k)=f(c_{k'})$.
	If $M\,\cdot\,M^\dagger\in\mathcal{CO}_{-1}$, then $r$ remains even unchanged.
	Therefore, let us define this minimal number $r$ of superimposed classical states as $r(\psi)$,
	\begin{eqnarray}
		r(\psi)=\inf\left\{r:|\psi\rangle=\sum_{k=1}^r \psi_k |c_k\rangle \wedge |c_k\rangle\in\mathcal C_0\right\}.
	\end{eqnarray}
	Obviously this number is 1, iff the state is an element of $\mathcal C_0$, and greater than one for a nonclassical pure state.

	Now let us consider a mixed state $\rho\in\mathcal Q$.
	This state can be written in various forms as a convex combination of pure states,
	\begin{eqnarray}\label{Eq:Dec1}
		\rho=\sum_i p_i|\psi_i\rangle\langle\psi_i|,
	\end{eqnarray}
	with $p_i>0$ and $\sum_i p_i=1$.
	In this case $\mu(\rho)$ can be obtained from a convex roof construction of $r(\psi)$~\cite{uhlmann}.
	In a particular decomposition given in Eq.~(\ref{Eq:Dec1}) the largest number of superposition of a pure state $|\psi_i\rangle$ can be found as $\sup_i\{r(\psi_i)\}$.
	Under all decompositions of $\rho$, the desired one is that with a minimum of needed superpositions.
	Thus, $\mu(\rho)$ is given by
	\begin{eqnarray}\label{eq:superpsoMeasure}
		\mu(\rho)=\inf \left\{\sup_i\{r(\psi_i)\}:\rho=\sum_i p_i|\psi_i\rangle\langle\psi_i|\right\}-1.
	\end{eqnarray}
	This number is 0, iff the mixed state is classical and greater than zero for nonclassical states.
	The number can become infinity, if no finite number of superpositions yields the given state.
	Let us highlight that states with an amount of quantumness up to $r$ define nested convex sets, $\mathcal C_{\mu\leq r}=\{\rho\in\mathcal Q:\mu(\rho)\leq r\}$ with $\mathcal C_r\subset \mathcal C_{r'}$ for $r\leq r'$.

	For convenience, it is also possible to map $\mu(\rho)$ together with a monotonically increasing function to another measure $\mu'$, e.g.,
	\begin{eqnarray}
		\mu'(\rho)= 1-\exp(-\mu(\rho))\in[0,1].
	\end{eqnarray}
	We also point out that the measures $\mu$ and $\mu'$ are invariant under classical invertible maps, $\mathcal{CO}_{-1}$, which is important for being compatible with the unambiguous ordering prescription.
	As we mentioned in Sec.~\ref{Sec:Motivation}, this is not true for a distance-based quantumness measure.
	Since $\mathcal{CO}_{-1}$ maps can be considered as a transformation of the underlying metric, a distance is in general not preserved.

	This function $\mu(\rho)$ in Eq.~(\ref{eq:superpsoMeasure}) is found to be an example of a quantumness measure based on convex ordering, which additionally characterizes the fundamental quantum superposition principle.
	In the case of coherent states it counts the minimal number of superpositions of (classical) coherent states needed to generate the state under study~\cite{UniQuant,Gehrke}.
	In the case of entanglement it represents the Schmidt number~\cite{SchmidtUni}.
	Hence, the given approach unifies and generalizes the previously considered methods.
	States with at most $r$ superpositions define nested, convex sets $\mathcal C_{\mu\leq r}$, which is advantageous for the construction of quantumness witnesses; cf.~\cite{MelWitness} and~\cite{SNWitness} for the construction of degree of nonclassicality witnesses and Schmidt number witnesses, respectively. 

	Let us note that the number of superpositions as a quantifier of quantumness in Eq.~(\ref{eq:superpsoMeasure}) may be further refined.
	For example the properties of the individual classical terms $|c_k\rangle$ in the superposition decomposition in Eq.~(\ref{Eq:SuperPosClassical}) could be taken into account.
	For certain practical applications, such as special quantum teleportation protocols, also the weighting coefficients $\psi_k$ can play a significant role.
	This, however, leads to operational quantumness measures, cf.~\cite{SchmidtUni}, which are important for quantifying the useful nonclassicality for particular applications.
	It might be also useful to use the purity of a quantum state $\rho$ to further refine quantumness measures.

\subsection{Example: Bits versus qubits}
	Another application of the superposition number is related to quantum information processing.
	A classical sequence of $N$ bits $\boldsymbol i=(i_1,\dots,i_N)$, with truth values ``0'' and ``1'', has a classical counterpart in a compound qubit quantum system $(\mathbb C^{2})^{\otimes N}$ as
	\begin{equation}
		|\boldsymbol i\rangle=|i_1\rangle\otimes\dots\otimes|i_N\rangle\in\mathcal C_0,
	\end{equation}
	where $|0\rangle$ and $|1\rangle$ are the ground and excited state, respectively, of any two-level system being described by the individual Hamiltonians
	\begin{equation}
		H=\frac{\hbar\omega}{2} \sigma_{z}, \mbox{ with } \sigma_z=|1\rangle\langle 1|-|0\rangle\langle 0|.
	\end{equation}
	Using classical probabilities, we only have statistical mixtures of sequences of bits as
	\begin{equation}
		\gamma=\sum_{\boldsymbol i\in\{0,1\}^N} p_{\boldsymbol i} |\boldsymbol i\rangle\langle \boldsymbol i|\in\mathcal C.
	\end{equation}
	Classical computational operations are those which compute -- including statistical imperfections or errors -- from a given classical sequence $\boldsymbol i$ another classical string $\boldsymbol j$ of $N$ bits with the probability $p(\boldsymbol j|\boldsymbol i)$:
	\begin{equation}
		\Lambda(|\boldsymbol i\rangle\langle \boldsymbol i|)=\sum_{\boldsymbol j\in\{0,1\}^N} p(\boldsymbol j|\boldsymbol i)\, |\boldsymbol j\rangle\langle \boldsymbol j|.
	\end{equation}
	An example of a classical invertible map is the $N$-bit NOT operation, $\Lambda(\,\cdot\,)={\rm NOT}^{\otimes N}(\,\cdot\,){\rm NOT}^{\otimes N}$, with ${\rm NOT}={\rm NOT}^\dagger=\sigma_x=|1\rangle\langle 0|+|0\rangle\langle 1|$.
	Please also note that the free unitary evolution with the given Hamiltonian also maps any classical string onto itself, see also~\cite{CMMV13}.

	Having identified the classical regime, we may study the quantum regime.
	Here, the pure states can be decomposed as
	\begin{equation}
		|\psi\rangle=\sum_{\boldsymbol i\in\{0,1\}^N}\psi_{\boldsymbol i}|\boldsymbol i\rangle,
	\end{equation}
	which is quantified by the superposition number
	\begin{equation}
		r(\psi)=|\{\psi_{\boldsymbol i}\neq0\}|,
	\end{equation}
	being the cardinality of the non-vanishing expansion coefficients $\psi_{\boldsymbol i}$.
	For example, a coherent superposition in a GHZ-type configuration, $(|0,\dots,0\rangle +|1,\dots,1\rangle)/\sqrt 2$, has a quantumness of $r=2$.
	This result, $r>1$, quantifies that such a state is beyond the classical information approach.
	A particular effect which can destroy these quantum interferences is given by decoherence, being the map
	\begin{equation*}
		\Lambda_{\rm dc}(\rho)=\int_{-\pi}^{+\pi} \!\!\!d\varphi\, p(\varphi) (\exp[i\varphi\sigma_z])^{\otimes N}\rho(\exp[-i\varphi\sigma_z])^{\otimes N},
	\end{equation*}
	for a classical phase distribution $p(\varphi)$.
	We observe that a full decoherence, i.e. a uniform distribution $p(\varphi)=1/(2\pi)$, maps any initial state onto the corresponding classical one,
	\begin{equation}
		\Lambda_{\rm dc}(|\psi\rangle\langle\psi|)=\sum_{\boldsymbol i\in\{0,1\}^N} |\psi_{\boldsymbol i}|^2|\boldsymbol i\rangle\langle \boldsymbol i|.
	\end{equation}
	Consistently our approach identifies that decoherence diminishes quantum properties.
	In the case of full decoherence we have $\mu(\Lambda_{\rm dc}(\rho))=0$ for any state $\rho\in\mathcal Q$, cf. Eq.~(\ref{eq:superpsoMeasure}).
	Therefore, our approach not only predicts an unambiguous order of quantumness in quantum information.
	It additionally characterizes the evolution of these quantum properties in realistic scenarios.

\section{Summary and conclusions}\label{Sec:SumCon}
	
	We have studied the quantification of quantum properties with a convex classical reference.
	It was outlined that distances-based measures, in general, lead to an ambiguous quantification.
	The origin of such a paradox lies in the fact that the nature of quantumness is an algebraic rather than a topological one:
	The mixture of classical states yields a convex subset of all quantum states.

	Based on the conservation of a classical feature under mixing, we have proposed a general convex ordering method.
	For handling classical processes or channels, we have additionally considered classical operations.
	We have shown that these transformations can be used to generalize our sorting procedure.
	By quantifying this order, we have obtained quantumness measures in a canonic form.
	In particular, quantumness probes based on the determination of quantum superpositions have been examined.
	The technique has been applied to typical examples in quantum physics such as entanglement and nonclassicality in terms of the Glauber-Sudarshan representation.
	Moreover, the embedding of classical information processing into the quantum domain led to a measure of the amount of quantumness in quantum information.
	In case of decoherence, we consistently retrieved the classical domain through our quantification.

	In conclusion, the number of quantum superpositions represents a vital measure to quantify the quantum nature of a system.
	Known examples have been considered in this context and they have been generalized.
	Ambiguities, as observed for other measures, do not occur and the role of reversible classical operations has been outlined.
	Our approach characterizes the quantum nature of states in terms of the fundamental superposition principle, and it naturally relates classical correlations to statistical mixing of states.
	We believe that this approach will be useful for characterizing even so-far unknown quantum effects in a broader context and for the general understanding of the strength of quantum effects in physical systems.

\appendix
\section{Normalization}\label{App:Normalization}
	Let us consider the normalization.
	It is more convenient to use the following sets,
	\begin{eqnarray}
		\mathcal Q'=&\{\lambda\rho: \lambda\geq0 \wedge \rho\in\mathcal Q \},\\
		\mathcal C'=&\{\lambda\rho: \lambda\geq0 \wedge \rho\in\mathcal C \},
	\end{eqnarray}
	instead of the normalized states, i.e. states with a unit trace: ${\rm tr}\,\rho=1$.
	The sets $\mathcal Q'$ and $\mathcal C'$ represent a cone construction over the sets $\mathcal Q$ and $\mathcal C$, respectively.
	According to these definitions, an element $\rho_3$ is element in $\mathcal C'$, if it can be written as a positive ($\lambda_1,\lambda_2\geq0$) linear combination of elements $\rho_1,\rho_2\in\mathcal C$,
	\begin{eqnarray}
		\rho_3=\lambda_1\rho_1+\lambda_2\rho_2.
	\end{eqnarray}
	In general, this linear combination is given by neither normalized states nor in a convex form.
	However, it can be rewritten in such a form.
	With ${\rm tr}\,\rho_3=\lambda_1{\rm tr}\,\rho_1+\lambda_2{\rm tr}\,\rho_2$, we obtain
	\begin{eqnarray*}
		\frac{\rho_3}{{\rm tr}\,\rho_3}{=}\frac{\lambda_1{\rm tr}\,\rho_1}{\lambda_1{\rm tr}\,\rho_1+\lambda_2{\rm tr}\,\rho_2} \frac{\rho_1}{{\rm tr}\,\rho_1}
		{+}\frac{\lambda_2{\rm tr}\,\rho_2}{\lambda_1{\rm tr}\,\rho_1+\lambda_2{\rm tr}\,\rho_2}\frac{\rho_2}{{\rm tr}\,\rho_2}.
	\end{eqnarray*}
	This is obviously a convex combination of normalized states.
	Therefore we can neglect without any loss of generality the normalization of the quantum states and perform the normalization at the end of our treatment.

\ack This work was supported by the Deutsche Forschungsgemeinschaft through SFB 652.
The authors gratefully acknowledge many stimulating discussions with Margarita and Vladimir Man'ko.


\section*{References}
\begin{thebibliography}{99}
	\bibitem{TraceDist} Hillery M 1987 {\it Phys. Rev.} A {\bf 35} 725
	\bibitem{TraceDist2} Hillery M 1989 {\it Phys. Rev.} A {\bf 39} 2994
	\bibitem{HSDist} Dodonov V V, Man'ko O V, Man'ko V I and W\"unsche A 1999 {\it Phys. Scr.} {\bf 59} 81
	\bibitem{HSDist2} Dodonov V V, Man'ko O V, Man'ko V I and W\"unsche A 2000 {\it J. Mod. Opt.} {\bf 47} 633
	\bibitem{BuresDist} Marian P, Marian T A and Scutaru H 2002 {\it Phys. Rev. Lett.} {\bf 88} 153601
	\bibitem{Gauss1} Lee C T 1991 {\it Phys. Rev.} A {\bf 44} R2775
	\bibitem{Gauss12} Lee C T 1995 {\it Phys. Rev.} A {\bf 52} 3374
	\bibitem{Gauss2} L\"utkenhaus N and Barnett S M 1995 {\it Phys. Rev.} A {\bf 51} 3340
	\bibitem{NegWiegner} Kenfack A and Zyczkowski K 2004 {\it J. Opt. B: Quantum Semiclass. Opt.} {\bf 6} 396
	\bibitem{FM10} Filippov S N and Man'ko V I 2010 {\it Phys. Scr.} {\bf T140} 014043 
	\bibitem{EntPot} Asboth J K, Calsamiglia J and Ritsch H 2005 {\it Phys. Rev. Lett.} {\bf 94} 173602
	\bibitem{AxiomEntM} Vedral V, Plenio M B, Rippin M A and Knight P L 1997 {\it Phys. Rev. Lett.} {\bf 78} 2275
	\bibitem{AxiomEntM2} Vidal G 2000 {\it J. Mod. Opt.} {\bf 47} 355
 	\bibitem{AxiomEntM3} Vedral V and Plenio M B 1998 {\it Phys. Rev.} A {\bf 57} 1619
 	\bibitem{SchmNummer} Sanpera A, Bru\ss{} D and Lewenstein M 2001 {\it Phys. Rev.} A {\bf 63} 050301(R)
 	\bibitem{TerhalSN} Terhal B M and Horodecki P 2000 {\it Phys. Rev.} A {\bf 61} 040301
	\bibitem{SchmidtUni} Sperling J and Vogel W 2011 {\it Phys. Scr.} {\bf 83} 045002
	\bibitem{UniQuant} Vogel W and Sperling J 2014 {\it Phys. Rev.} A {\bf 89} 052302
        \bibitem{Eisert-QC} Gross D, Flammia S T and Eisert J 2009 {\it Phys. Rev. Lett.} {\bf 102} 190501
	\bibitem{Gehrke} Gehrke C, Sperling J and Vogel W 2012 {\it Phys. Rev.} A {\bf 86} 052118
	\bibitem{Fischer} Hall M J W 2000 {\it Phys. Rev.} A {\bf 62} 012107
	\bibitem{FKMMSV10} Facchi P, Kulkarni R, Man'ko V I, Marmo G, Sudarshan E C G and Ventriglia F 2010 {\it Phys. Lett.} A {\bf 374} 4801 
	\bibitem{EntropyInformation} Vedral V 2002 {\it Rev. Mod. Phys.} {\bf 74} 197
	\bibitem{Context} Grudka A, Horodecki K, Horodecki M, Horodecki P, Horodecki R, Joshi P, K\l{}obus W and W\'{o}jcik A 2014 {\it Phys. Rev. Lett.} {\bf 112} 120401
	\bibitem{MM09} Man'ko M A and Man'ko V I 2009 {\it Theor. Math. Phys.} {\bf 160} 995 
	\bibitem{MM14a}  Man'ko V I and Markovich L A 2014 {\it J. Russ. Laser Res.} {\bf 35} 355 
	\bibitem{MM14} Man'ko M A and Man'ko V I 2014 {\it J. Phys.: Conf. Ser.} {\bf 538} 012016 
	\bibitem{DiscordReview} Modi K, Brodutch A, Cable H, Paterek T and Vedral V 2012 {\it Rev. Mod. Phys.} {\bf 84} 1655
	\bibitem{RMP-Horo} Horodecki R, Horodecki P, Horodecki M, and Horodecki K 2009 {\it Rev. Mod. Phys.} {\bf 81} 865
	\bibitem{GSRep2}  Sudarshan E C G 1963 {\it Phys. Rev. Lett.} {\bf 10} 277
	\bibitem{GSRep1}  Glauber R J 1963 {\it Phys. Rev.} {\bf 131} 2766
	\bibitem{EntRep} Sperling J and Vogel W 2009 {\it Phys. Rev. A} {\bf 79} 042337
	\bibitem{OpSumRep} Kraus K 1983 {\it States, Effects and Operations, Fundamental Notions of Quantum Theory} (Berlin: Academic)
	\bibitem{uhlmann} Uhlmann A 1998 {\it Open Syst. Inf. Dyn.} {\bf 5} 209
	\bibitem{MelWitness} Mraz M, Sperling J, Vogel W and Hage B 2014 {\it Phys. Rev.} A {\bf 90} 033812
	\bibitem{SNWitness} Sperling J and Vogel W 2011 {\it Phys. Rev.} A {\bf 83} 042315
	\bibitem{CMMV13} Chr\'{s}ci\'{n}ski D, Man'ko V I, Marmo G and Ventriglia F 2013 {\it Phys. Scr.} {\bf 87} 045015 

\end{thebibliography}
\end{document}